\documentclass[twocolumn]{cinc}
\usepackage{graphicx}
\usepackage{algorithm}
\usepackage{algorithmic}
\usepackage[table,xcdraw,dvipsnames]{xcolor}
\usepackage{subcaption}
\usepackage[acronym]{glossaries}
\usepackage[per-mode=symbol]{siunitx}
\usepackage{url}

\makeglossaries 

\newacronym{psd}{PSD}{power spectral density}
\newacronym{hr}{HR}{heart rate}
\newacronym{fft}{FFT}{Fast Fourier Transform}
\newacronym{ma}{MA}{motion artifacts}
\newacronym{dtw}{DTW}{dynamic time warping}
\newacronym{pdf}{PDF}{probability density function}
\newacronym{ppe}{PPE}{posterior probability estimation}
\newacronym{acc}{ACC}{accelerometer}
\newacronym{mae}{MAE}{mean absolute error}
\newacronym{dl}{DL}{deep learning}

\begin{document}
\bibliographystyle{cinc}

\title{Robust PPG-exclusive Heart Rate Estimation \\ During Maximal Exercise Using Population-Based Priors}


\author {Elisabetta De Giovanni$^{1}$, Iñigo Urteaga$^{1,2}$, Matteo Croci$^{1,2,3}$, Tomas Teijeiro$^{1,3}$ 
\\
\ \\ 
 $^1$ Basque Center for Applied Mathematics (BCAM), Bilbao, Spain \\
 $^2$ Basque Foundation for Science (Ikerbasque), Bilbao, Spain. \\
 $^3$ University of the Basque Country (EHU), Leioa, Spain}

\maketitle

\begin{abstract}
    

    PPG is the main technology for wearable heart rate (HR) monitoring. However, motion artifacts during intense physical activity reduce its reliability due to spectral overlap with the cardiac signal. This study investigates whether severely contaminated PPG signals can yield reliable HR estimates without auxiliary sensors by leveraging prior knowledge of HR dynamics. We acquired simultaneous PPG and ECG signals from 21 subjects during a maximal cycloergometer test. Using Leave-One-Subject-Out cross-validation, we built HR evolution priors from the ground-truth ECG and applied Dynamic Time Warping (DTW) to align normalized HR curves and estimate a barycenter average. 
    DTW aligns normalized HR curves with minimal error across all subjects, validating robust population-based priors. 
    For each test subject, we combined the PPG Power Spectral Density (PSD) with the prior distribution to generate a ``posterior PSD’’. We identified and ranked candidate HR trajectories using heuristic search within the ``posterior PSD'' and selected the most likely path. Our approach reduced \gls{mae} in BPM relative to the ground-truth ECG by a median of 63.3\% (IQR: 47.3\%-74.6\%) compared to a PPG+accelerometer benchmark. These results show that physiological priors and DTW-based alignment enable robust, PPG-exclusive HR monitoring during high-intensity exercise.
\end{abstract}

\section{Introduction} 


PPG remains a key technology for non-invasive wearable \gls{hr} monitoring, due to its low cost and user comfort. However, during intense physical activity, \gls{ma} frequently corrupt the cardiac signal, compromising the reliability of the \gls{hr} estimation. 

This study investigates whether heavily contaminated PPG signals can still yield reliable HR estimations without auxiliary sensors, such as \gls{acc}, by leveraging prior knowledge of HR dynamics. 

To date, proposed methods to remove  \gls{ma} for continuous \gls{hr} estimation fall into two main categories~\cite{Sicbaldi2025}: classical (model-driven) and machine learning (data-driven) algorithms. Classical approaches involve motion artifact correction, spectrum estimation, and spectral peak tracking. The machine learning algorithms,  including \gls{dl}, use supervised non-linear regression to learn features from time- or frequency-domain data without physical assumptions. Within the \gls{dl} frameworks, Kechris et al.~\cite{Kechris2025} recently proposed \texttt{KID-PPG}, which integrates physiological priors and artifact handling into PPG-based \gls{hr} estimation by combining data-driven inference with domain expertise. Bieri et al.~\cite{Bieri2023} propose a \gls{dl} framework (\texttt{BeliefPPG}) with two approaches, PPG+\gls{acc} and PPG-only. \texttt{BeliefPPG} models \gls{hr} dynamics using belief propagation over a hidden Markov model, capturing predictive uncertainty. Additionally, hybrid methods combine both approaches, as the work of Guo et al.~\cite{Guo2026}, which uses support vector machines to classify signal quality as low or high before applying spectral analysis. According to the benchmarks of Sicbaldi et al.~\cite{Sicbaldi2025}, \gls{dl} algorithms consistently outperform classical ones, which tend to underestimate \gls{hr}, with \texttt{BeliefPPG}~\cite{Bieri2023} being the most robust. However, most of these works rely on \gls{acc} data to remove \gls{ma}.

For this reason, we present a PPG-exclusive \gls{hr} estimation method that uses population-based priors for high-intensity exercise. We validate our method on in-house datasets collected during maximal and submaximal cycloergometer testing. 
We select \texttt{BeliefPPG} as the reference to compare our proposed method. Finally, we highlight the range of applicability and limitations of our proposed approach. 

%





\section{PPG-exclusive HR estimation}

\begin{figure}[ht]
\includegraphics[width=0.99\linewidth]{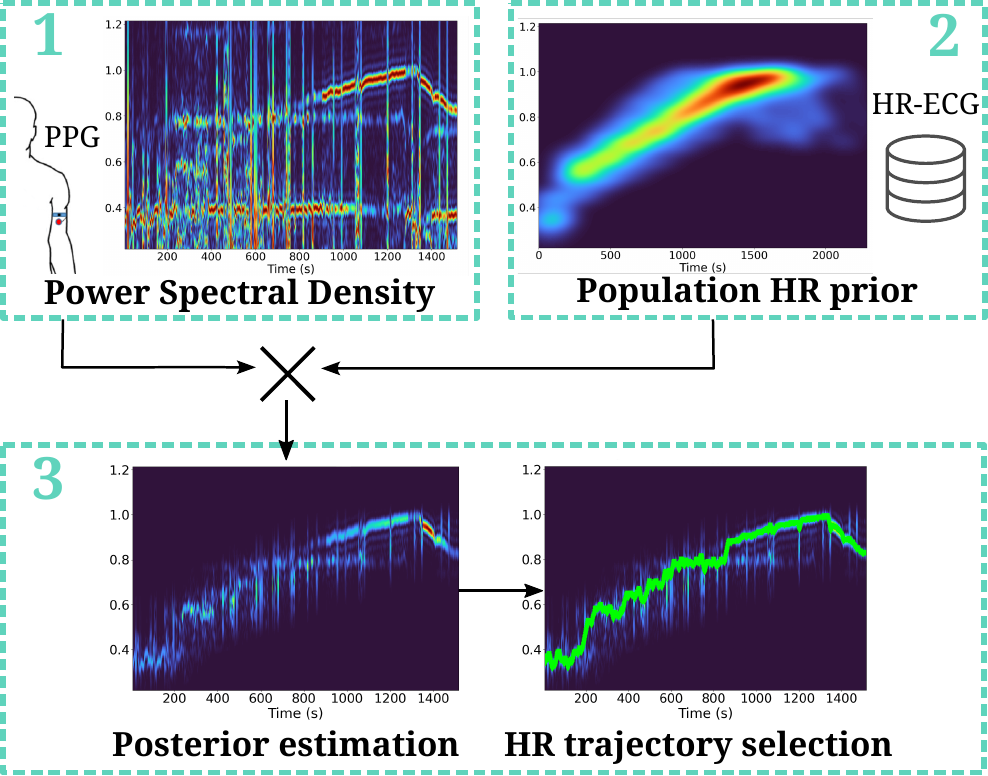}
\caption{Overview of our proposed PPG-exclusive \gls{hr} estimation using population-based priors. These specific plots correspond to an example subject of the main dataset (Sub7). Step 1 shows the PSD of the PPG over time, which is highly corrupted in this subject. Step 2 shows the prior computed on the other subjects of the dataset. Step 3 shows the posterior estimation and the selected \gls{hr} trajectory.}
\label{fig:block-diagram}
\end{figure}

Our proposed method, named \texttt{PPG-Prior}, includes three main steps as shown in Figure~\ref{fig:block-diagram}. The first step computes the \gls{psd} of the PPG signal over time. The second step involves building the \gls{hr} dynamics prior over a population of subjects different from the one analyzed. The final step combines the observations (\gls{psd}) and the prior to obtain a posterior approximation, analyzes suitable \gls{hr} trajectory candidates, and selects the optimal one. The full code of \texttt{PPG-Prior} is available as open-source software\footnote{\scriptsize{\url{https://github.com/elidg/ppg-prior-hr-estimation}}}.

\subsection{Power spectral density over time}

As a first step of our proposed method, the raw PPG signal is bandpass filtered ([0.5, 10]~\SI{}{\hertz}) and detrended. Using an 8-second window with a 1-second slide~\cite{DeGiovanni2016}, we compute the online \gls{psd} via periodogram estimation after applying a real-input inverse \gls{fft} bandpass filter. Each normalized window is restricted to the physiological HR frequency range ([40, 220]~BPM). As shown in Figure~\ref{fig:block-diagram} (step 1), motion artifacts affect the resulting \gls{psd} through transient bursts, represented by vertical lines, and periodic noise traces, represented by horizontal lines.

\subsection{Population-based prior}

To model \gls{hr} dynamics during maximal exercise, the second step computes a population-based prior from ECG-derived RR intervals in a reference dataset. Each series is smoothed using a 30-beat rolling window and normalized to the subject's peak \gls{hr}. This collection serves two purposes: deriving a representative trend by computing a barycenter average to align the candidate trajectories, and generating a prior distribution. For the first purpose, we applied \gls{dtw} to align the normalized HR series and quantify morphological similarities
across subjects. Specifically, we use the \texttt{soft-DTW}~\cite{Cuturi2017}  method, which regularizes standard \gls{dtw} by computing a soft minimum over all alignment costs, making it well-suited for averaging time series. We apply a Leave-One-Subject-Out cross-validation methodology to build the HR evolution prior from the training population. We construct the prior distribution as the \gls{pdf} of the \gls{hr} series collection through a multivariate kernel density estimation, with the same grid (normalized frequency, time) as each PPG \gls{psd}. 

\subsection{Posterior HR estimation} \label{sec:ppe}

\begin{algorithm}[H]
\caption{\gls{hr} Trajectory Extraction and Selection}
\label{alg:traj-sel}
\begin{algorithmic}[1]
\REQUIRE $PSD$, $PDF$, limits $\epsilon_{\text{local}}$ and $\epsilon_{\text{rmse}}$ \label{line:ppe}

\STATE $PPE \gets PSD \times PDF$
\STATE $\mathcal{T} \gets \emptyset$

\WHILE{length$(\mathcal{T})\le$ 50} \label{line:searchS}
    \STATE $\mathbf{y} \gets \text{ViterbiDecode}(PPE, \epsilon_{\text{local}})$
    \IF{$\mathbf{y}$ is valid \AND $\forall \mathbf{a} \in \mathcal{T}, \text{RMSE}(\mathbf{y}, \mathbf{a}) \ge \epsilon_{\text{rmse}}$}
    \STATE $S(\mathbf{y}) \gets \sum_{t=1}^{T} \log PPE[y_t, t]$ 
    \STATE $\mathcal{T} \gets \mathcal{T} \cup \{(\mathbf{y}, S(\mathbf{y}))\}$
    \ENDIF
\ENDWHILE \label{line:searchE}

    \STATE $\mathbf{y}^*\gets\text{SelectTopCandidate}(\mathcal{T})$ \COMMENT{max $S(\mathbf{y})$ or min DTW distance to barycenter} 
    \label{line:top}

\RETURN $\mathbf{y}^*$
\end{algorithmic}
\end{algorithm}
The final step of our method combines the PPG-derived \gls{psd} with the population-based prior to compute the per-subject \gls{ppe} and extract the optimal \gls{hr} trajectory (Algorithm~\ref{alg:traj-sel}). We generate the \gls{ppe} via multiplication of the \gls{psd} and the prior \gls{pdf} on a normalized \gls{hr} over a fixed time grid ([40, 220]~BPM, 0.92~BPM resolution, 1-second time steps). To extract candidate \gls{hr} trajectories ($\mathcal{T}$) from the \gls{ppe}, we implement an iterative Viterbi search that decodes optimal sequences while constraining local \gls{hr} variations ($\epsilon_{\text{local}}$) to enforce physiological plausibility (between one time window and the next). After each iteration, the $\epsilon_{\text{rmse}}$ parameter discards trajectories that are too similar to already selected candidates, ensuring diversity among the identified paths. We explore two scoring metrics to select the final \gls{hr} trajectory candidate: 1) the algorithm sorts the candidate trajectories by their cumulative log-score over time on \gls{ppe} ($S(\mathbf{y})$) and selects the top trajectory (\textit{Top-Traj}). 
2) the algorithm minimizes the \gls{dtw} distance between candidate trajectories and the \texttt{soft-DTW} barycenter (\textit{Min-DTW}). 
The latter strategy is particularly useful when the posterior remains highly corrupted, requiring greater reliance on the prior, i.e., on the learned population-based physiological knowledge.

 

 

 



\section{Experimental setup and results} 

We analyzed data from healthy subjects engaged in regular physical activity who performed maximal and submaximal cycloergometer tests with varying increment durations and power loads. The \textsc{Max-HIE} dataset includes ECG, PPG, and 3-axis \gls{acc} data from 21 subjects (14 males, 7 females, ages 20–37; 3-min, 30-watt increments), extending the maximal exercise dataset in~\cite{DeGiovanni2021}. PPG and \gls{acc} data were recorded from the arm (repositioned if needed), and single-lead chest ECG (leads I–II) served as ground truth, with all signals resampled at 250 Hz. For prospective validation and generalization assessment, we collected submaximal test data (\textsc{Submax-HIE}) from five volunteers (4 females, 1 male, ages 26–35) using a Polar Verity Sense (PPG: 176 Hz; \gls{acc}: 416 Hz), Hexoskin ProShirt (HR every 1 s), or Shimmer ECG (RR-derived HR) if needed, with 2-min increments. Experimental protocols were approved by CER-VD (ref. 2016-00308) and CEIm-E (ref. PI2025217).
We compared our \texttt{PPG-Prior} method to the \texttt{BeliefPPG} algorithm~\cite{Bieri2023}, which uses PPG with or without \gls{acc} data and reports every 2 seconds. We report the \gls{mae} calculated second-by-second against ECG and the percentage \gls{mae} drop versus \texttt{BeliefPPG}. For \textsc{Max-HIE}, we used a Leave-One-Subject-Out prior; for \textsc{Submax-HIE}, we used priors from (i) \textsc{Max-HIE} and (ii) the Physionet \textsc{ACTES}~\cite{PhysioNet-actes-cycloergometer-exercise-1.0.0} dataset (1-min, 15-watt increments).

\subsection{Maximal exercise results}

\begin{table}[ht]
\caption{Summary results for maximal exercise test of \textsc{Max-HIE} dataset}
\label{tbl:result-maximal}
\resizebox{\linewidth}{!}{%
\begin{tabular}{l|c|c|c|c}
 &
  \multicolumn{3}{c|}{\textbf{MAE (BPM) w.r.t HR-ECG}} &
  \multicolumn{1}{c}{} \\ \cline{2-4}
\textbf{Metric} &
  \multicolumn{1}{c|}{\textbf{\begin{tabular}[c]{@{}l@{}}BeliefPPG \\ (PPG+ACC)\end{tabular}}} &
  \multicolumn{1}{c|}{\textbf{\begin{tabular}[c]{@{}l@{}}BeliefPPG \\ (only PPG) \end{tabular}}} &
  \textbf{\begin{tabular}[c]{@{}l@{}}PPG-Prior \\ \textit{Top-Traj} \end{tabular}} &
  \textbf{\begin{tabular}[c]{@{}l@{}}MAE drop (\%) \\ w.r.t. BeliefPPG \\ (PPG+ACC)\end{tabular}}\\ \hline
\textbf{Min} &
  \multicolumn{1}{c|}{4.53} &
  \multicolumn{1}{c|}{{\textcolor{red}{20.76}}} &
  \multicolumn{1}{c|}{1.74} &
  \multicolumn{1}{c}{6.3\%} \\
\textbf{Max} &
  \multicolumn{1}{c|}{60.46} &
  \multicolumn{1}{c|}{\textcolor{red}{68.83}} &
  \multicolumn{1}{c|}{23.42} &
  \multicolumn{1}{c}{85.4\%} \\
\textbf{Median} &
  \multicolumn{1}{c|}{21.35} &
  \multicolumn{1}{c|}{{\textcolor{red}{47.49}}} &
  \multicolumn{1}{c|}{8.88} &
  \multicolumn{1}{c}{\textbf{63.3\%}} \\
\textbf{Q1} &
  \multicolumn{1}{c|}{15.61} &
  \multicolumn{1}{c|}{{\textcolor{red}{32.74}}} &
  \multicolumn{1}{c|}{2.92} &
  \multicolumn{1}{c}{\textbf{47.3\%}} \\
\textbf{Q3} &
  \multicolumn{1}{c|}{28.11} &
  \multicolumn{1}{c|}{{\textcolor{red}{56.90}}} &
  \multicolumn{1}{c|}{11.94} &
  \multicolumn{1}{c}{\textbf{74.6\%}}
\end{tabular}%
}
\end{table}

Table~\ref{tbl:result-maximal} summarizes the main results of our \texttt{PPG-Prior}, highlighting a median \gls{mae} drop of 63.3\% relative to \texttt{BeliefPPG} with PPG+ACC, and up to 85.4\%. We also reported the \gls{mae} of \texttt{BeliefPPG} using only PPG. However, we do not include a direct comparison, as it consistently underperforms relative to PPG+ACC across all subjects. Across the full dataset, our proposed framework using the \textit{Top-Traj} scoring method achieved a \gls{mae} under 3 BPM in 28.6\% of cases and under 5 BPM in 38.1\% of cases, substantially outperforming \texttt{BeliefPPG}, which never reduces the MAE below 3 BPM and only reduces it below 5 BPM in 14.3\% of cases.
\begin{figure}[t]
\includegraphics[width=\linewidth]{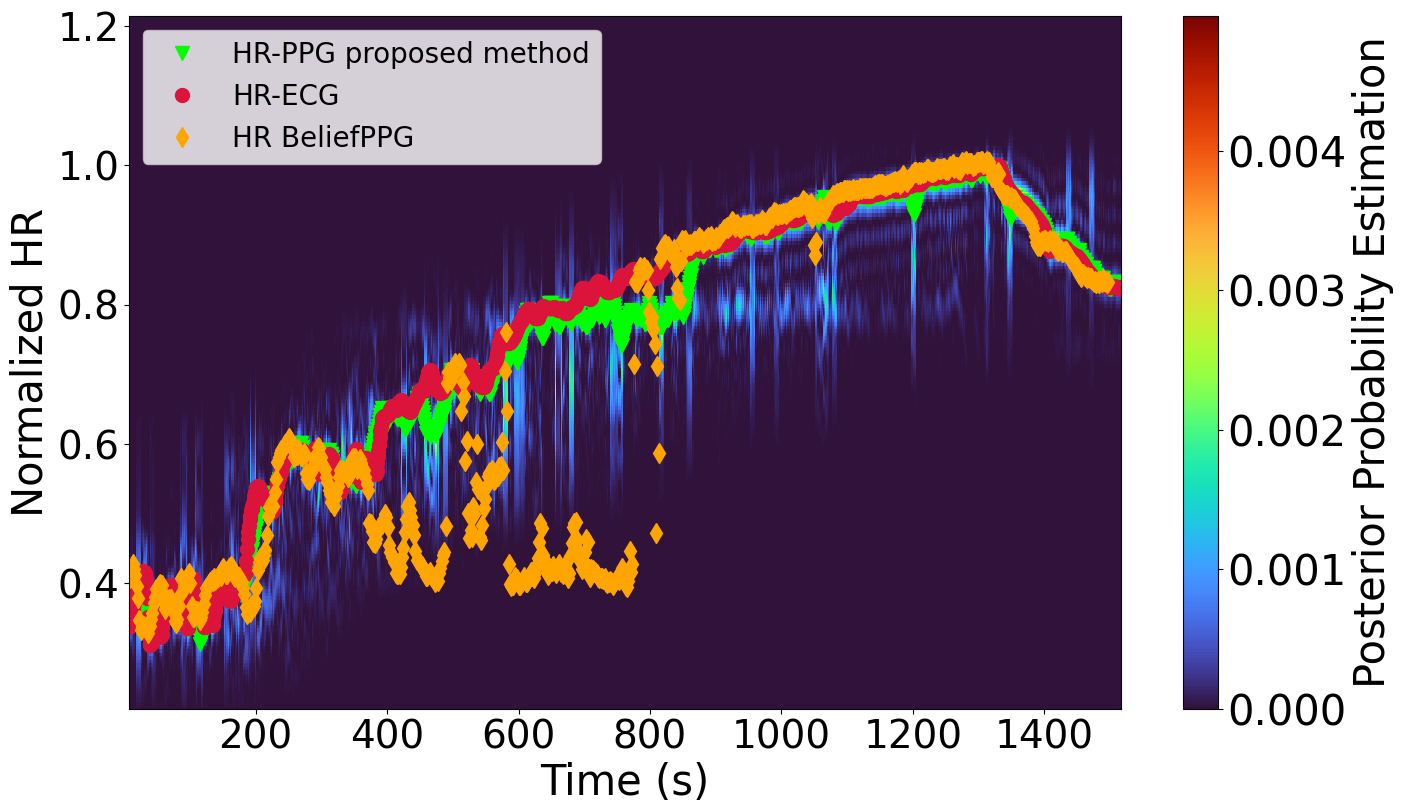}
\caption{Normalized \gls{hr} trajectory selected on \gls{ppe} by our proposed method for Sub7. Comparison with ground-truth ECG and \texttt{BeliefPPG} with PPG+ACC data.}
\label{fig:ppe-hr}
\end{figure}

\begin{table}[ht]
\caption{Per-subject results for \textsc{Max-HIE} dataset in example subjects reporting two path-selection methods}
\label{tbl:result-max-subs}
\resizebox{\linewidth}{!}{%
\begin{tabular}{l|c|c|c|c|c}
& 
  \multicolumn{3}{c|}{\textbf{MAE (BPM) w.r.t HR-ECG}} &
  \multicolumn{2}{c}{\textbf{\begin{tabular}[c]{@{}l@{}}MAE drop (\%) \\ w.r.t. BeliefPPG \end{tabular}}} \\ \cline{2-6}
\textbf{Subject} &
  \multicolumn{1}{c|}{\textbf{\begin{tabular}[c]{@{}l@{}} BeliefPPG \\ (PPG+ACC)\end{tabular}}} &
  \multicolumn{1}{c|}{\textbf{\begin{tabular}[c]{@{}l@{}}PPG-Prior \\ \textit{Min-DTW} \end{tabular}}} &
  \multicolumn{1}{c|}{\textbf{\begin{tabular}[c]{@{}l@{}}PPG-Prior \\ \textit{Top-Traj} \end{tabular}}} &
  \multicolumn{1}{c|}{\textbf{\begin{tabular}[c]{@{}l@{}}PPG-Prior\\ \textit{Min-DTW} \end{tabular}}} & 
  \textbf{\begin{tabular}[c]{@{}l@{}}PPG-Prior\\ \textit{Top-Traj}\end{tabular}} \\ \hline
\textbf{Sub7} &
  \multicolumn{1}{c|}{{16.10}} &
  \multicolumn{1}{c|}{{3.69}} &
  \multicolumn{1}{c|}{{\textbf{2.92}}} &
  \multicolumn{1}{c|}{{77.1\%}} &
  {\textbf{81.9\%}} \\ 
\textbf{Sub21} &
  \multicolumn{1}{c|}{4.62} &
  \multicolumn{1}{c|}{3.67} &
  \multicolumn{1}{c|}{{\textbf{1.74}}} &
  \multicolumn{1}{c|}{20.6\%} &
  {\textbf{62.3\%}} \\ 
\textbf{Sub5} &
  \multicolumn{1}{c|}{{36.74}} &
  \multicolumn{1}{c|}{\textcolor{Green}{\textbf{14.55}}} &
  \multicolumn{1}{c|}{\textcolor{orange}{\textbf{20.09}}} &
  \multicolumn{1}{c|}{\textcolor{Green}{\textbf{60.4\%}}} &
  \textcolor{orange}{\textbf{45.3\%}} \\ 
\textbf{Sub10} &
  \multicolumn{1}{c|}{46.85} &
  \multicolumn{1}{c|}{\textcolor{Green}{\textbf{10.94}}} &
  \multicolumn{1}{c|}{\textcolor{orange}{\textbf{13.40}}} &
  \multicolumn{1}{c|}{\textcolor{Green}{\textbf{76.6\%}}} & 				
  \textcolor{orange}{\textbf{71.4\%}} 
\end{tabular}%
}
\end{table}
Table~\ref{tbl:result-max-subs} reports the results for example subjects to highlight specific cases and the range of applicability of our proposed method. In this case, we evaluate the performance of both scoring methods described in Section~\ref{sec:ppe}. The results highlight the reliance on the prior when the posterior remained severely corrupted, such as in Sub5 and Sub10. Sub7 achieved the best outcome, even though its \gls{psd} was highly corrupted, as shown in Figure~\ref{fig:block-diagram}, Step 1.  Figure~\ref{fig:ppe-hr} illustrates how our proposed method outperforms \texttt{BeliefPPG} in Sub7, especially in the most corrupted segment. Even in low-noise scenarios (Sub21), our method achieves a 62.3\% drop in \gls{mae}. Moreover, in approximately 33\% of the dataset, the \textit{Min-DTW} scoring method improves the MAE by up to 5.5 BPM. 
We remark that the MAE remains high in the cases considered in Table 2 due to \gls{hr} trend gaps in the original \gls{psd} observations. More advanced modeling approaches are thus needed to address these complexities.

\subsection{Submaximal exercise results} 

Table~\ref{tbl:result-submax} shows the results for the \textsc{Submax-HIE} data. In four out of five cases, \texttt{BeliefPPG} performed reasonably well (MAE $<$ 6 BPM), while our proposed method (with \textit{Top-Traj}) decreased the MAE by up to 49.3\% using the \textsc{Max-HIE} prior in three out of five cases. In two cases, our approach yielded higher errors than \texttt{BeliefPPG}. However, if we use the \textsc{ACTES} prior, which has a different increment duration and power load, the MAE is reduced by up to 91.2\%. This is particularly evident for A002, which had a highly corrupted PPG and a short test (only three 2-min increments). This result shows that incorporating prior knowledge of the specific exercise protocol is crucial for accurate \gls{hr} estimation, highlighting the need to integrate exercise-context awareness in future developments. 

\begin{table}[ht]
\caption{Per-subject results for \textsc{Submax-HIE} in healthy volunteers (A) and elite basketball players (B), using different priors}
\label{tbl:result-submax}
\resizebox{\linewidth}{!}{%
\begin{tabular}{l|c|c|c|c|c}
& 
  \multicolumn{3}{c|}{\textbf{MAE (BPM) w.r.t HR-ECG}} &
  \multicolumn{2}{c}{\textbf{\begin{tabular}[c]{@{}l@{}}MAE drop (\%) w.r.t. BeliefPPG \end{tabular}}} \\ \cline{2-6}
\textbf{Subject} &
  \multicolumn{1}{c|}{\textbf{\begin{tabular}[c||]{@{}l@{}}BeliefPPG \\ (PPG+ACC)\end{tabular}}} &
  \multicolumn{1}{c|}{\textbf{\begin{tabular}[c]{@{}l@{}}PPG-Prior \\ with \textsc{Max-HIE}\end{tabular}}} &
  \multicolumn{1}{c|}{\textbf{\begin{tabular}[c]{@{}l@{}}PPG-Prior \\ with \textsc{ACTES}\end{tabular}}} &
  \multicolumn{1}{c|}{\textbf{\begin{tabular}[c]{@{}l@{}}PPG-Prior \\with \textsc{Max-HIE} \end{tabular}}} &
  \multicolumn{1}{c}{\textbf{\begin{tabular}[c]{@{}l@{}}PPG-Prior \\ with \textsc{ACTES} \end{tabular}}} \\ \hline
\textbf{A000} &
  3.60 &
  \textcolor{red}{\textbf{5.71}} &
  \textcolor{Green}{\textbf{2.68}} &
  \textcolor{red}{\textbf{-58.6\%}} &
  \textcolor{Green}{\textbf{25.5\%}} \\
\textbf{A001} &
  4.04 &
  3.71 &
  3.13 &
  \textbf{8.2\%} &
  \textbf{22.5\%} \\
\textbf{A002} &
  15.56 &
  \textcolor{red}{\textbf{39.23}} &
  \textcolor{Green}{\textbf{1.38}} &
  \textcolor{red}{\textbf{-152.0\%}} &
  \textcolor{Green}{\textbf{91.2\%}} \\
\textbf{B001} &
  3.75 &
  1.98 &
  1.63 &
  {\textbf{47.1\%}} &
  \textbf{56.4\%} \\
\textbf{B023} &
  5.64 &
  2.86 &
  2.87 &
  \textbf{49.3\%} &
  \textbf{49.1\%}
\end{tabular}%
}
\end{table}








\section{Limitations and future work} 

Despite its promising results, our proposed method has multiple limitations that highlight key directions for future work. First, the current trajectory selection lacks a segment-level adaptive strategy. Future work should develop an automated method to evaluate posterior reliability and dynamically select the top candidate posterior trajectories over prior-based paths, particularly in noisy segments. Second, our validation is currently restricted to incremental maximal and submaximal exercise protocols. We plan to expand testing to non-incremental protocolized tasks during real-world sports training sessions. Furthermore, prior empirical learning is restricted to specific available datasets, which may miss relevant data gaps in the estimation that require physiological trend modeling and forecasting to fill. Finally, a key long-term goal is to develop personalized priors for tailored continuous monitoring.


\section{Conclusion}

We presented a new method for PPG-exclusive \gls{hr} estimation using population-based priors validated on maximal and submaximal exercise tests. Our approach achieved a 63.3\% median \gls{mae} reduction compared to a state-of-the-art method requiring \gls{acc} data. The results highlight how incorporating exercise-specific prior knowledge effectively reduces \gls{ma} without auxiliary motion sensors. Future work will integrate physiological trend forecasting and adaptive path selection, tailored to broader, non-incremental exercise contexts. 

\balance



 

\section*{Acknowledgments}  

E. De Giovanni was supported by the EU Horizon Europe programme under the Marie Sklodowska-Curie actions with grant agreement 101202439 (SWEATHEART).
This work has also been supported by the grants PID2024-156173OA-I00 (PREDISENSE), RYC2023-045922-I, \allowbreak RYC2022-036312-I, and CEX2021-001142-S, funded by MICIU/AEI/10.13039/501100011033 and ESF+. I. Urteaga, M. Croci and T. Teijeiro were partially supported by the Basque Government through the BERC 2022-2025 program. 
The authors also acknowledge Prof. G. P. Millet, Dr. N. Bourdillon from ISSUL, Bilbao Kirolak, and the volunteers for their availability in the data collection.

\bibliography{refs}


  
  
      

\begin{correspondence}
Elisabetta De Giovanni
\\
Alameda de Mazarredo 14, 48009 Bilbao, Spain\\
edegiovanni@bcamath.org
\end{correspondence}

\end{document}